\documentclass{PoS}

\title{HAWC Observation of Supernova Remnants and Pulsar Wind Nebulae}

\ShortTitle{HAWC Observations of SNRs and PWNe}

\author{\speaker{C. M. Hui}$^a$, and H. Zhou$^a$ for the HAWC Collaboration$^b$\\
        \llap{$^a$}Department of Physics, Michigan Technological University\\
        1400 Townsend Drive, Houghton, MI, USA\\
        \llap{$^b$}For a complete author list, see \href{http://www.hawc-observatory.org/collaboration/icrc2015.php}{www.hawc-observatory.org/collaboration/icrc2015.php}\\
        E-mail: \email{cmhui@mtu.edu}, \email{hzhou1@mtu.edu}}

\abstract{
The majority of Galactic TeV gamma-ray sources are pulsar wind nebulae (PWNe) and supernova remnants (SNRs), and the most common association for unidentified sources is PWN. Many of these sources were discovered in TeV by imaging air Cherenkov telescopes using overlapping pointed observations over sections of the Galactic plane. The HAWC observatory is a survey type instrument in the Northern hemisphere with an energy range of 100 GeV to 100 TeV. Preliminary analysis of data recorded with the partially completed HAWC array taken since 2013 shows extended detections that are coincident with known TeV SNRs and PWNe. The full array became operational in early 2015 and has been steadily surveying the Northern sky since. I will discuss detections in HAWC data taken since 2013 associated with PWNe and SNRs.
}

\FullConference{The 34th International Cosmic Ray Conference,\\
		30 July- 6 August, 2015\\
		The Hague, The Netherlands}

\begin{document}

\section{Introduction}
% need to reword this paragraph to differ from paper.
The most commonly identified TeV Galactic sources are supernova remnants (SNRs) and pulsar wind nebulae (PWNe).  SNRs are postulated as cosmic-ray acceleration site since they can accelerate particles via diffusive shock acceleration and provide the amount of energy needed to produce the measured local cosmic-ray energy density.  In PWNe, electrons effectively gain energy at the termination shock where the pulsar wind is terminated by the surrounding gas, emitting TeV gamma rays via inverse Compton scattering.

Many of the gamma-ray sources in the Galactic plane are in fact unidentified due to their extended nature enveloping multiple sources identified in other wavelengths.  Morphological and spectral studies are crucial for associating with other wavelength observations and for distinguishing leptonic and hadronic gamma-ray production processes.

Over 60 of the known TeV sources were discovered during the H.E.S.S. survey of the southern sky.  Similar surveys have been performed by Milagro \cite{milagroSurvey} and ARGO \cite{argoSurvey} in the northern sky, along with targeted observations by imaging air Cherenkov telescopes, resulting in over 10 discoveries.  HAWC is a recently inaugurated TeV gamma-ray surveying instrument located in Mexico with unprecedented sensitivity at $>10$\,TeV.  The high energy reach is useful in determining spectral cutoffs that could be indicative of the acceleration process, and in studying energy-dependent morphology by combining with lower energy experiments.  In this proceeding, we present a survey of the sources in the Galactic plane using data from the partially built HAWC detector.

%  The region surveyed with the partial HAWC array contains locations of 73 sources from the Fermi 3FGL catalog, 47 of which are without known astronomical associations. only inner galaxy region though.  Mention how many TeV sources also are within the detection regions (inner galaxy and cygnus?)

\section{The HAWC Gamma-Ray Observatory}
The High Altitude Water Cherenkov (HAWC) observatory is a second generation TeV gamma-ray detector based on the water Cherenkov technique developed by the Milagro gamma-ray observatory \cite{milagro}.  It is located at Sierra Negra, Mexico, at an elevation of 4100\,m.  The construction of the array is complete, and the detector was inaugurated in March 2015.  It consists of 300 water Cherenkov detectors (WCDs), covering an area of $\sim22,000\,\textrm{m}^2$.  Each WCD is 7.3\,m in diameter and 4.5\,m in depth, containing $\sim200,000\,\textrm{L}$ of purified water.  Within each WCD are three 8'' PMTs and a central 10'' PMT facing upwards to detect the Cherenkov radiation from charged particles produced in extensive air shower initiated by a primary gamma ray or cosmic ray.  The HAWC observatory operates at $>95\%$ duty cycle with a 2\,sr instantaneous field of view.  For a complete description of the experiment, please see \cite{Smith-ICRC}.

\subsection{Data and Analysis}
Since the summer of 2013, the HAWC observatory has been collecting data with a partially built array.  In this proceeding, $\sim280$ live days of data taken with 1/3 of the detector (HAWC-111) are presented.
%Between August 2013 and July 2014, the HAWC-111 detector has a trigger rate of 15\,kHz.  
The core and angle reconstruction of the triggered events are optimized for gamma rays.  Therefore, gamma-induced showers can be distinguished from hadronic showers using the goodness of fit.  An additional parameter used for gamma-hadron discrimination is the maximum charge found at a distance $>40$\,m from the core as a function of the size of the air shower event.  The gamma-hadron separation cuts are optimized on the Crab Nebula for this dataset.

After gamma-hadron selection cuts, the background is estimated by the direct integration method, which convolves the local coordinates and event distribution as a function of time \cite{milagro}.  For the maps presented in this proceeding, the average time integration window is 2 hours.  The maps are smoothed with a double Gaussian point spread function derived from data on the Crab Nebula and significance is calculated using equation 17 from Li \& Ma \cite{lima}.  The median energy of the HAWC-111 dataset after gamma-hadron separation cuts is $\sim 10\,$TeV and higher depending on the declination.  The performance of the HAWC-111 array has been evaluated using observations of gamma rays from the Crab Nebula; please see \cite{Paco-ICRC} for details.

\begin{figure}
  \centering
  \includegraphics[width=0.95\textwidth]{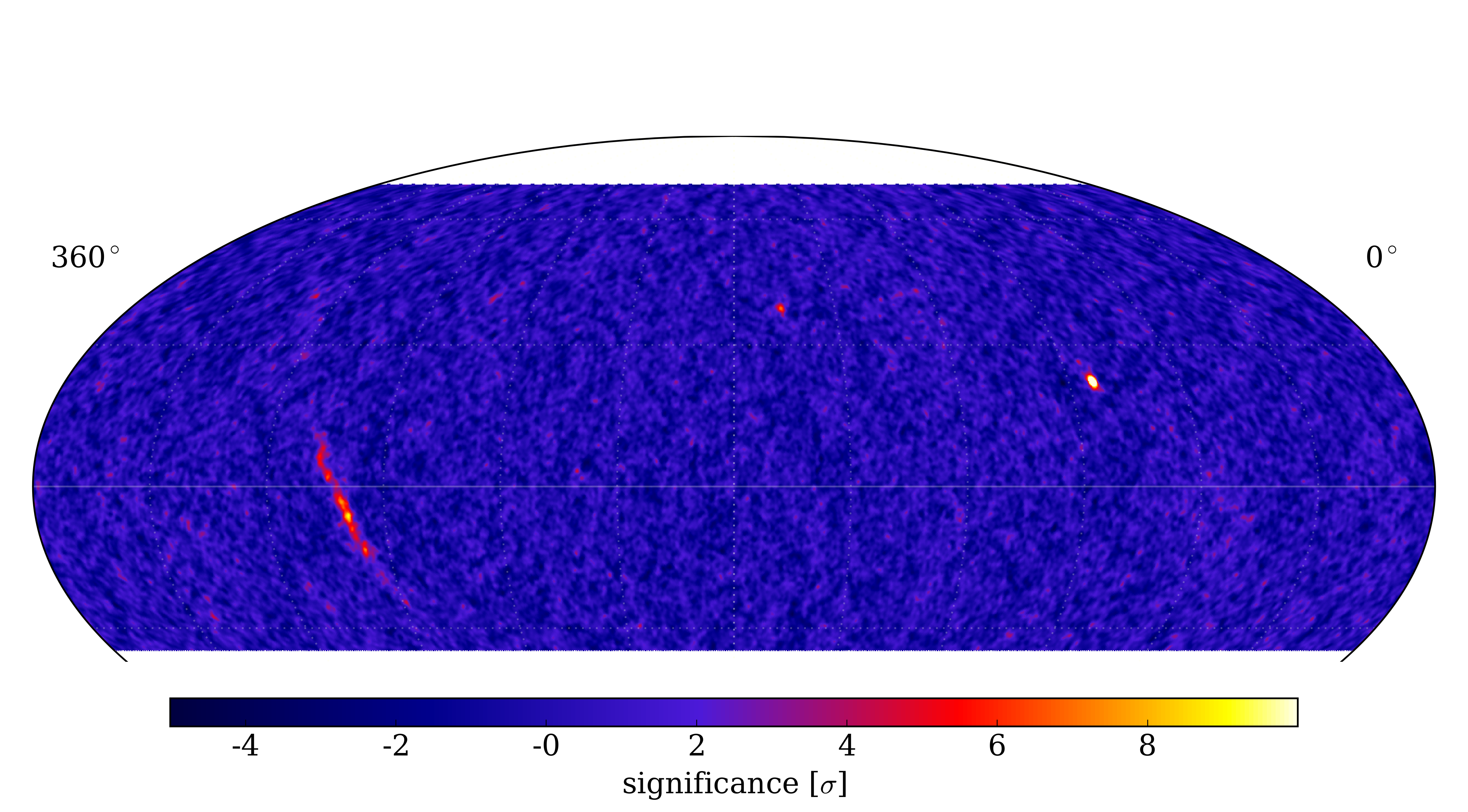}
  \caption{Skymap of 280 days of data with 1/3 of the HAWC array.  The analysis is optimized on the Crab Nebula, which is detected at $>24\,\sigma$.  The Galactic plane is clearly visible at $>5\,\sigma$.} 
  \label{fig:skymap}
\end{figure}

Figure \ref{fig:skymap} shows the skymap of the HAWC-111 dataset with an analysis optimized on the Crab Nebula.  The Crab Nebula is detected at $>24\,\sigma$ along with extended regions of $>5\,\sigma$ along the Galactic plane.  

% total integration hour is 4310 -> 180 days?
% HAWC250 is 3566 -> 148.5 days
To identify sources and study their spectra and morphologies, a maximum likelihood analysis to test different physics models is currently being developed and is presented in \cite{likelihood}.  Using the likelihood framework, the Crab Nebula is detected at with a test statistic ($TS$) of 491.7 (equivalent of $22.2\,\sigma$) at right ascension $\alpha=83.53^\circ \pm 0.06^\circ_{stat}$ and declination $\delta=22.06^\circ \pm 0.06^\circ_{stat}$.  The flux, assuming a simple power law with spectral index of 2.60, is $(4.62 \pm 0.23_{stat})\times 10^{-13}\,\textrm{TeV}^{-1} \textrm{cm}^{-2} \textrm{s}^{-1}$ at 5\,TeV (energy where the dependence of index assumption is minimized for the flux normalization).  The flux are compatible with previous TeV measurements to within 15\%  \cite{hessCrab,magicCrab,verCrab}.
% Crab flux at 5TeV (HAWC111): 4.62E-13 2.23E-14 2.60 0.10
% our integral flux >1TeV is 1.9 pm 0.1? e-11
% our flux at 1TeV is 3.0e-11
% Whipple at 1TeV (3.25 pm 0.14)e-11, index 2.49, integral flux >1TeV (2.1 pm 0.2)e-11
%% at 5 TeV 5.91e-13, ours is 78% of Whipple
% HEGRA at 1TeV (2.79 pm 0.02)e-11, index 2.59, integral flux >1TeV 1.75e-11
%% at 5 TeV 4.32e-13, ours is 107% of HEGRA
% HESS PL at 1TeV (3.45 pm 0.05)e-11, index 2.63, integral flux >1TeV is 2.26 pm 0.08 e-11
%% at 5 TeV 5.01e-13, ours is 92% of HESS PL
% MAGIC at 300GeV (5.7 pm 0.2)e-10, index 2.48, integral flux >200GeV 1.96 pm 0.05 e-10
%% at 5 TeV 5.32e-13, ours is 87% of MAGIC PL
% VERITAS at 1TeV (3.48 pm 0.14)e-11, index 2.65
%% at 5 TeV 4.89e-13, ours is 94% of VERITAS

\section{Galactic Gamma-Ray Sources}
\begin{figure}
  \centering
  \includegraphics[width=\textwidth]{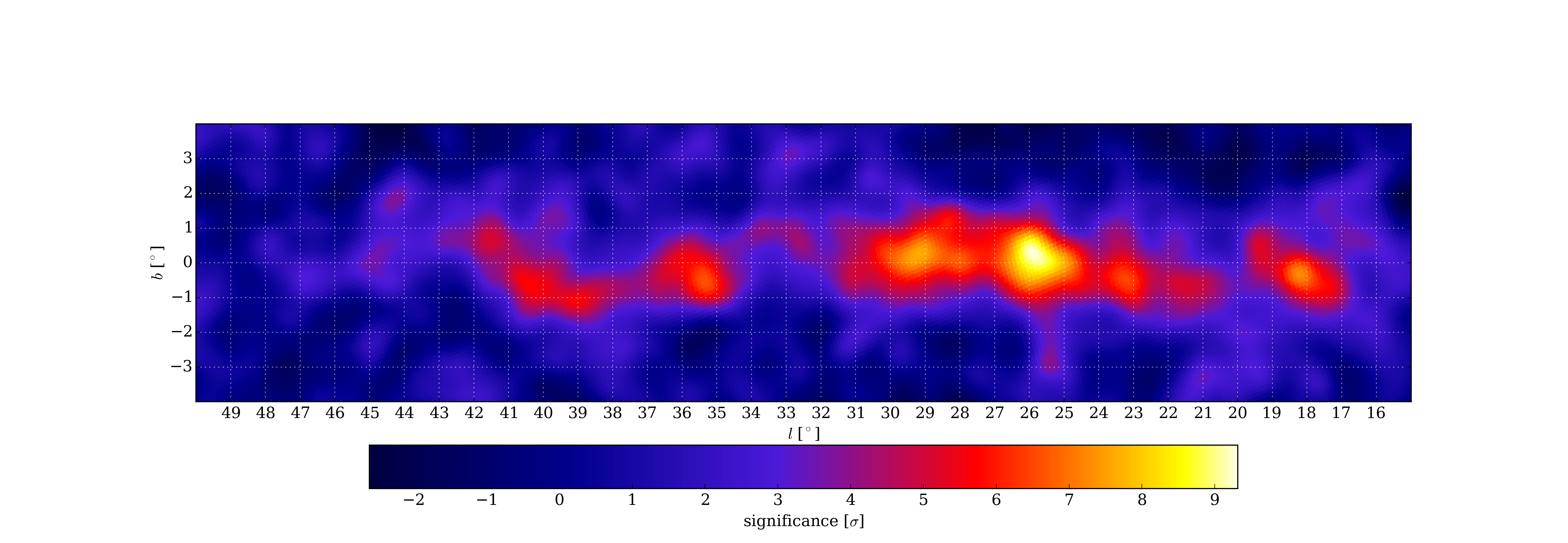}
  \caption{Significance map of the inner Galactic plane.} 
  \label{fig:IGzoom}
\end{figure}

% brief descripion of the detections and the source classes in the HAWC data
Within $\pm4.0^\circ$ of the Galactic plane, there are extended areas with pixels $>5\,\sigma$ pre-trials.  Figure \ref{fig:IGzoom} shows the significance map of this region.  The likelihood analysis is applied to the region of interest within Galactic longitude $15^\circ < l < 50^\circ$ and latitude $-4^\circ < b < 4^\circ$.  Source candidates are identified by three parameters: $\alpha$, $\delta$, and flux with a spectral index assumption of 2.3.  An index assumption of 2.3 is used because it is representative of those measured in Galactic TeV sources.  The region is modeled by an iterative process in which new point sources are added to the model until the change in $TS$ is $<15$ (about $3\sigma$) with the addition of another source.  Eleven seed sources are identified by this process, most are coincident with known TeV sources and will be discussed in an upcoming publication.  In this proceeding, we report on the five most significant source candidates as listed in table \ref{tab:src}.  All the position coordinates reported below are in J2000 epoch and with statistical uncertainties.

% use the TS map from likelihood analysis

\begin{table}
\centering
\caption{List of five most significant source candidates in the inner Galactic region of the HAWC-111 dataset.}
\label{tab:src}
\begin{tabular}{|c|c|c|c|c|c|}
\hline
source candidate& $\alpha$ & $\delta$ & pre-trial  & possible & type \\
 & & & significance & association & \\ \hline
1HWC\,J1857+023 & $284.3^\circ \pm 0.2^\circ$ & $2.3^\circ \pm 0.2^\circ$ & $7.1\,\sigma$ & HESS\,J1857+026, & PWN \\
 & & & & HESS\,J1858+020 & UID \\\hline
1HWC\,J1838-060 & $279.6^\circ \pm 0.3^\circ$ & $-6.0^\circ \pm 0.2^\circ$ & $7.0\,\sigma$ & HESS\,J1841-055 & UID \\\hline
1HWC\,J1825-133 & $276.3^\circ \pm 0.1^\circ$ & $-13.3^\circ \pm 0.2^\circ$ & $6.4\,\sigma$ & HESS\,J1825-137 & PWN \\\hline
1HWC\,J1844-031 & $281.0^\circ \pm 0.2^\circ$ & $-3.1^\circ \pm 0.2^\circ$ & $5.8\,\sigma$ & HESS\,J1843-033 & UID  \\\hline
1HWC\,J1907+062 & $286.8^\circ \pm 0.2^\circ$ & $6.2^\circ \pm 0.2^\circ$ & $5.7\,\sigma$ & MGRO\,J1908+06 & UID \\ \hline
\end{tabular}
\end{table}

\subsection{Supernova Remnants}
Within the inner Galaxy region described above, there are three known TeV SNRs: W51, W49B, and G$015.4+00.1$.  Their reported TeV flux are on the order of 1\% of the Crab.  The HAWC-111 dataset is not sensitive to sources with this flux level in the inner Galaxy.
% W51, 3% of Crab
% W49B, <1% of Crab
% G015.4+0.1, 2% of Crab

\subsection{Pulsar Wind Nebulae}
There are seven identified TeV PWNe within the inner Galaxy region.  Two of the HAWC detections are near a TeV PWN.  A search on pulsed emission from pulsars is presented in \cite{Cesar-ICRC}.

1HWC\,J1825-133 has a pre-trials significance of $6.4\,\sigma$ at $\alpha=276.3^\circ \pm 0.1^\circ$ and $\delta=-13.3^\circ \pm 0.2^\circ$.  It is $\sim 0.5^\circ$ away from the centroid position of HESS\,J1825-137, which is an extended PWN.  
%The integral flux above 1\,TeV is 47\% of the Crab Nebula flux, which is $\sim 2/3$ of the reported flux from H.E.S.S. \cite{hess1825}.  
The simple power law spectrum was reported to be unlikely by \cite{hess1825} and more detailed spectral study will be done with data from the full HAWC array.

1HWC\,J1857+023 is $7.1\,\sigma$ pre-trials at $\alpha=284.3^\circ \pm 0.2^\circ$, $\delta=2.3^\circ \pm 0.2^\circ$.  It is $\sim 0.4^\circ$ away from HESS\,J1857+026 and HESS\,J1858+020, a PWN identified by MAGIC \cite{magic1857} and an unidentified source \cite{hessuid} respectively.  We are unable to resolve the two sources using the HAWC-111 dataset. 
%The integral flux above 1\,TeV is 30\% of the Crab Nebula flux and is compatible with the H.E.S.S. and MAGIC measurements of the PWN \cite{hessuid,magic1857}.

% HESS J1825-137, 17% of Crab
% HESS J1837-069, 13% of Crab
% MAGIC J1857.2+0263,  % HESS J1857+026
%%%
% HESS J1831-098, 4% of Crab, 3 sigma
% HESS J1846-029, 2% of Crab, 6.2e-13, -2.26
% IGR J18490-0000, 1.5% of Crab
% HESS J1912+101, 10% of Crab (3 sigma blobs)

\subsection{Unidentified Sources}
1HWC\,J1844-031 has a pre-trials significance of $5.8\,\sigma$ at $\alpha=281.0^\circ \pm 0.2^\circ$ and $\delta=-3.1^\circ \pm 0.2^\circ$.  It is spatially coincident with the unidentified source HESS\,J1843-033 \cite{hess1843} and extends towards the PWN HESS\,J1846-029.  %The integral flux above 1\,TeV is 30\% of the Crab Nebula flux, there is no previously reported flux at this location.

1HWC\,J1838-060 is $7.0\,\sigma$ pre trials at $\alpha=279.6^\circ \pm 0.3^\circ$ and $\delta=-6.0^\circ \pm 0.2^\circ$.%, with an integral flux above 1\,TeV at 41\% of the Crab Nebula flux.  
This detection is located in the middle of HESS\,J1837-069 and HESS\,J1841-055.  The ARGO-YBJ collaboration reported a $5.3\,\sigma$ excess at a location consistent with the HAWC detection, but attributed it to HESS\,J1841-055, which has a $0.41^\circ \times 0.25^\circ$ extent reported by H.E.S.S. \cite{argo1841,hessuid}.
%The flux reported by \cite{argo1841} is higher than the flux reported here by a factor of 3.

1HWC\,J1907+062 is $5.7\,\sigma$ pre trials at $\alpha=286.8^\circ \pm 0.2^\circ$ and $\delta=6.2^\circ \pm 0.2^\circ$, which is consistent with previously reported positions of MGRO\,J1908+06.  %The integral flux above 1\,TeV is 20\% of the Crab Nebula flux, and is consistent with previous reported flux by H.E.S.S. and VERITAS.

% HESS J1858+020 -- mentioned in PWN section for HESS J1857+023
% MGROJ1908+06
% HESS J1848-018, massive star cluster
% HESS J1843-033
% HESS J1841-055
% HESS J1834-087, 
% HESS J1832-093, 1% of Crab, 2.6 index (3-4 sigma blob)
% HESS J1831-098, 4% of Crab 2.1 index (3-4 sigma blob)
% LS5039, 3% of Crab, binary

\section{Outlook}
A preliminary TeV survey of the northern sky is presented using data taken with a partially built HAWC array.  Using a maximum likelihood analysis, eleven point-like seed sources are modeled in the inner Galactic region of $15^\circ < l < 50^\circ$ and latitude $-4^\circ < b < 4^\circ$.  The five most significant source candidates and their possible counterparts are presented here.  The detailed results of the HAWC-111 dataset will be discussed in an upcoming publication.

% mention Cygnus region and why we did not include it here.  At the time of this proceedings, the analysis of the Cyngus region is still ongoing...

The HAWC observatory is now completed and is continuously taking data and monitoring the gamma-ray sky.  A multiwavelength study using a joint likelihood framework to combine Fermi-LAT and HAWC data is currently being developed \cite{threeML}.  Joint fits of Galactic sources using data from Fermi-LAT and HAWC are presented in \cite{Hao-ICRC}.

\section*{Acknowledgments}
\footnotesize{
We acknowledge the support from: the US National Science Foundation (NSF);
the US Department of Energy Office of High-Energy Physics;
the Laboratory Directed Research and Development (LDRD) program of
Los Alamos National Laboratory; Consejo Nacional de Ciencia y Tecnolog\'{\i}a (CONACyT),
Mexico (grants 260378, 55155, 105666, 122331, 132197, 167281, 167733);
Red de F\'{\i}sica de Altas Energ\'{\i}as, Mexico;
DGAPA-UNAM (grants IG100414-3, IN108713,  IN121309, IN115409, IN111315);
VIEP-BUAP (grant 161-EXC-2011);
the University of Wisconsin Alumni Research Foundation;
the Institute of Geophysics, Planetary Physics, and Signatures at Los Alamos National Laboratory;
the Luc Binette Foundation UNAM Postdoctoral Fellowship program.
}

\bibliography{icrc2015-0323}

\providecommand{\href}[2]{#2}\begingroup\raggedright\begin{thebibliography}{10}

\bibitem{milagroSurvey}
A.~A. {Abdo} et~al. {\em ApJL} {\bf 664} (Aug., 2007) L91--L94.

\bibitem{argoSurvey}
B.~{Bartoli} et~al. {\em ApJ} {\bf 779} (Dec., 2013) 27.

\bibitem{milagro}
R.~{Atkins} et~al. {\em ApJ} {\bf 595} (Oct., 2003) 803--811.

\bibitem{Smith-ICRC}
{\bf HAWC} Collaboration, A.~Smith, {\it {HAWC: Design, Operation,
  Reconstruction and Analysis}},  in {\em Proc. 34th ICRC}, (The Hague, The
  Netherlands), August, 2015.

\bibitem{lima}
T.-P. {Li} and Y.-Q. {Ma} {\em ApJ} {\bf 272} (Sept., 1983) 317--324.

\bibitem{Paco-ICRC}
{\bf HAWC} Collaboration, F.~Salesa, {\it {Observations of the Crab Nebula with
  Early HAWC Data}},  in {\em Proc. 34th ICRC}, (The Hague, The Netherlands),
  August, 2015.

\bibitem{likelihood}
{\bf HAWC} Collaboration, P.~Younk, {\it {A high-level analysis framework for
  HAWC}},  in {\em Proc. 34th ICRC}, (The Hague, The Netherlands), August,
  2015.

\bibitem{hessCrab}
F.~{Aharonian} et~al. {\em A\&A} {\bf 457} (Oct., 2006) 899--915.

\bibitem{magicCrab}
J.~{Albert} et~al. {\em ApJ} {\bf 674} (Feb., 2008) 1037--1055.

\bibitem{verCrab}
E.~{Aliu} et~al. {\em ApJL} {\bf 781} (Jan., 2014) L11.

\bibitem{Cesar-ICRC}
{\bf HAWC} Collaboration, C.~Alvarez, {\it {Searching for Very High Energy
  Emission from Pulsars Using the High Altitude Water Cherenkov (HAWC)
  Observatory}},  in {\em Proc. 34th ICRC}, (The Hague, The Netherlands),
  August, 2015.

\bibitem{hess1825}
F.~{Aharonian} et~al. {\em A\&A} {\bf 460} (Dec., 2006) 365--374.

\bibitem{magic1857}
J.~{Aleksi{\'c}} et~al. {\em A\&A} {\bf 571} (Nov., 2014) A96.

\bibitem{hessuid}
F.~{Aharonian} et~al. {\em A\&A} {\bf 477} (Jan., 2008) 353--363.

\bibitem{hess1843}
S.~{Hoppe}, {\it {The H.E.S.S. survey of the inner Galactic plane}},  {\em
  International Cosmic Ray Conference} {\bf 2} (2008) 579--582,
  [\href{http://arxiv.org/abs/0710.3528}{{\tt arXiv:0710.3528}}].

\bibitem{argo1841}
B.~{Bartoli} et~al. {\em ApJ} {\bf 767} (Apr., 2013) 99.

\bibitem{threeML}
G.~Vianello, {\it {The Multi-Mission Maximum Likelihood framework}},  in {\em
  Proc. 34th ICRC}, (The Hague, The Netherlands), August, 2015.

\bibitem{Hao-ICRC}
{\bf HAWC} Collaboration, H.~Zhou, {\it {TeV Observations of the Galactic Plane
  with HAWC and Joint Analysis of GeV Data from Fermi}},  in {\em Proc. 34th
  ICRC}, (The Hague, The Netherlands), August, 2015.

\end{thebibliography}\endgroup

\end{document}